# Implementing Push-Pull Efficiently in GraphBLAS

Carl Yang
University of California, Davis
Lawrence Berkeley National Laboratory
ctcyang@ucdavis.edu

Aydın Buluç
Lawrence Berkeley National Laboratory
abuluc@lbl.gov

John D. Owens
University of California, Davis
jowens@ece.ucdavis.edu

## ABSTRACT

We factor Beamer's push-pull, also known as direction-optimized breadth-first-search (DOBFS) into 3 separable optimizations, and analyze them for generalizability, asymptotic speedup, and contribution to overall speedup. We demonstrate that masking is critical for high performance and can be generalized to all graph algorithms where the sparsity pattern of the output is known *a priori*. We show that these graph algorithm optimizations, which together constitute DOBFS, can be neatly and separably described using linear algebra and can be expressed in the GraphBLAS linear-algebra-based framework. We provide experimental evidence that with these optimizations, a DOBFS expressed in a linear-algebra-based graph framework attains competitive performance with state-of-the-art graph frameworks on the GPU and on a multi-threaded CPU, achieving 101 GTEPS on a Scale 22 RMAT graph.

## 1 INTRODUCTION

Push-pull, also known as direction-optimized breadth-first-search (DOBFS), is a key optimization for making breadth-first-search (BFS) run efficiently [7]. According to the Graph Algorithm Platform [6], no fewer than 32 out of the top 37 entries on the Graph500 benchmark (a suite for ranking the fastest graph frameworks in the world) use direction-optimizing BFS. Since its discovery, it has been extended to other traversal-based algorithms [10, 29]. One of our contributions in this paper is factoring Beamer's direction-optimized BFS into 3 separable optimizations, and analyzing them independently—both theoretically and empirically—to determine their contribution to the overall speed-up. This allows us to generalize these optimizations to other graph algorithms, as well as fit it neatly into a linear algebra-based graph framework. These 3 optimizations are, in increasing order of specificity:

(1) Change of direction: Use the *push* direction to take advantage of knowledge that the frontier is small, which we term *input sparsity*. When the frontier becomes large, go back to the *pull* direction.
(2) Masking: In the *pull* direction, there is an asymptotic speed-up if we know *a priori* the subset of vertices to be updated, which we term *output sparsity*.
(3) Early-exit: In the *pull* direction, once a single parent has been found, the computation for that undiscovered node ought to exit early from the search.

GraphBLAS is an effort by the graph analytics community to formulate graph algorithms as sparse linear algebra [12]. The goal of the GraphBLAS API specification is to outline the common, high-level operations such as vector-vector inner product, matrix-vector product, matrix-matrix product, and define the standard interface for scientists to use these functions in a hardware-agnostic manner. This way, the runtime of the GraphBLAS implementation can make the difficult decisions about optimizing each of the GraphBLAS operations on a given piece of hardware.

Previous work by Beamer et al. [8] and Besta et al. [10] have observed that push and pull correspond to column- and row-based matvec (Optimization 1). However, this realization has not made it into the sole GraphBLAS implementation in existence so far, namely SuiteSparse GraphBLAS [15]. In SuiteSparse GraphBLAS, the BFS executes in only the forward (push) direction.

The key distinction between our work and that of Shun and Blelloch [29], Besta et al. [10], and Beamer et al. [8] is that while they take advantage of *input sparsity* using change of direction (Optimization 1), they do not analyze using *output sparsity* through masking (Optimization 2), which we show theoretically and empirically (in Table 1 and 2 respectively) is critical for high performance. Furthermore, we submit this speed-up extends to all algorithms for which there is *a priori* information regarding the sparsity pattern of the output such as triangle counting and enumeration [3], adaptive PageRank [20], batched betweenness centrality [12], maximal independent set [13], and convolutional neural networks [14].

Since the input vector can be either sparse or dense, we refrain from referring to this operation as SpMSpV (sparse matrix-sparse vector) or SpMV (sparse matrix-dense vector). Instead, we will refer to it as matvec (short for matrix-vector multiplication and known in GraphBLAS as GrB_mxv). Our contributions in this paper are:

(1) We provide theoretical and empirical evidence of the asymptotic speed-up from masking, and show it is proportional to the fraction of nonzeroes in the expected output, which we term *output sparsity*.

(2) We provide empirical evidence that masking is a key optimization required for BFS to attain state-of-the-art performance on GPUs.
(3) We generalize the concept of masking to work on all algorithms where *output sparsity* is known before the operation.
(4) We show that direction-optimized BFS can be implemented in GraphBLAS with minimal change to the interface by virtue of an isomorphism between push-pull and column- and row-based matvec.

## 2 BACKGROUND & PRELIMINARIES

### 2.1 Breadth-first-search

We consider breadth-first search on a directed or undirected graph $G = (V, E)$. $V$ is the set of vertices of $G$, and $E$ is the set of all ordered pairs $(u, v)$, with $u, v \in V$ such that $u$ and $v$ are connected by an edge in $G$. A graph is undirected if for all $v, u \in V : (v, u) \in E \iff (u, v) \in E$. Otherwise, it is directed. For directed graphs, a vertex $u$ is the child of another vertex $v$ if $(v, u) \in E$ and the parent of another vertex $v$ if $(u, v) \in E$.

Given a source vertex $s \in V$, a BFS is a full exploration of graph $G$ that produces a spanning tree of the graph, containing all the edges that can be reached from $s$, and the shortest path from $s$ to each one of them. We define the depth of a vertex as the number of hops it takes to reach this vertex from the root in the spanning tree. The visit proceeds in steps, examining one BFS level at a time. It uses three sets of vertices to keep track of the state of the visit: the *frontier* contains the vertices that are being explored at the current depth, *next* has the vertices that can be reached from *frontier*, and *visited* has the vertices reached so far.

### 2.2 Direction-optimized breadth-first-search

Push is the standard textbook way of thinking about BFS. At the start of each push step, each vertex in the *frontier* looks for its children and adds them to the *next* set if they have not been visited before. Once all children of the current frontier have been found, the discovered children are added to the visited array with the current depth, the depth is incremented, and the *next* set becomes the *frontier* of the next BFS step.

Pull is an alternative algorithmic formulation of BFS, yielding the same results but computing the *next* set in a different way. At the start of each pull step, each vertex in the *unvisited* set of vertices looks for its parents. If at least one parent is part of the *frontier*, we include the vertex in the *next* set.

Because either push or pull is a valid option to compute each step, we can achieve better overall BFS performance if we make the optimal algorithmic choice at each step. This is the key idea behind direction-optimized breadth-first-search (DOBFS), also known as push-pull BFS [7]. Push-pull can also be used for other traversal-based algorithms [10, 29]. DOBFS implementations use a heuristic function after each step to determine whether push or pull will be more efficient on the next step.

### 2.3 Notation

We use the MATLAB colon notation where $\mathbf{A}(:, i)$ denotes the $i$th column, $\mathbf{A}(i, :)$ denotes the $i$th row, and $\mathbf{A}(i, j)$ denotes the element at the $(i, j)$th position of matrix $\mathbf{A}$. We use $.*$ for the elementwise multiplication operator. For two frontiers $\mathbf{u}, \mathbf{v}$, their elementwise multiplication product $\mathbf{w} = \mathbf{u} .* \mathbf{v}$ is defined as $\mathbf{w}(i) = \mathbf{u}(i) * \mathbf{v}(i) \; \forall i$. The number of nonzero elements in matrix $\mathbf{A}$ is $nnz(\mathbf{A})$ and the number of nonzero elements of vector $\mathbf{v}$ is $nnz(\mathbf{v})$.

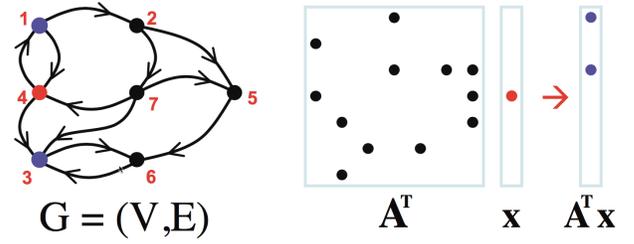

**Figure 1: Matrix-graph duality. The adjacency matrix A is the dual of graph $G$. The matvec is the dual of the BFS graph traversal. Figure is based on Kepner and Gilbert's book [21].**

For a set of nodes $\mathbf{v}$, we will say the number of outgoing edges $nnz(m_\mathbf{v}^+)$ is the sum of the number of outgoing edges of all nodes that belong to this set. Outgoing edges are denoted by a superscript '+', and incoming edges are denoted by a superscript '−'. That is, the number of incoming edges for a set of nodes $\mathbf{v}$ is,

$$nnz(m_\mathbf{v}^-) = \sum_{i: \, \mathbf{v}(i) \neq 0} nnz(\mathbf{A}^T(i, :)). \quad (1)$$

### 2.4 Traversal is matvec

Since the early days of graph theory, the duality between graphs and matrices has been established by the popular representation of a graph as an adjacency matrix [24]. It has become well-known that a vector-matrix multiply in which the matrix represents the adjacency matrix of a graph is equivalent to one iteration of breadth-first-search traversal. This is shown in Figure 1.

## 3 TYPES OF MATVEC

The next sections will make a distinction between the different ways the matvec $\mathbf{y} \leftarrow \mathbf{A}\mathbf{x}$ can be computed. We define matvec as the multiplication of a sparse matrix with a vector on the right. This definition allows us to classify algorithms as row-based and column-based without ambiguity. We distinguish between SpMV (sparse matrix-dense vector multiplication) and SpMSpV (sparse matrix-sparse vector multiplication). Our analysis differs from previous work that focuses on SpMV, while we concentrate on SpMSpV. Our novelty also comes from analysis of their masked variants, which is a mathematical formalism for taking advantage of *output sparsity* and to the best of our knowledge does not exist in the literature. It is worth noting that the sparse vectors are assumed to be implemented as sorted lists of indices and values.

Henceforth, we will refer to SpMV as row-based matvec, and SpMSpV as column-based matvec. This is justified because although it is possible to implement SpMV in a column-based way and SpMSpV in a row-based way, it is generally more efficient to implement SpMV by iterating over rows of the matrix [30] and SpMSpV by fetching columns of the matrix $\mathbf{A}(:, i)$ for which $\mathbf{x}(i) \neq 0$ [2]. Here, we are talking about SpMV and SpMSpV without direct dependence on graph traversal. We use the common, untransposed problem description $\mathbf{y} \leftarrow \mathbf{A}\mathbf{x}$ instead of that specific to graph traversal case.

## 3.1 Row- and column-based matvec

We wish to understand, from a matrix point of view, which of row- and column-based matvec is more efficient. We quantify efficiency with the random-access memory (RAM) model of computation. Since we assume the input vector must be read in both row- and column-based matvec, we will focus our attention on the number of random memory accesses into matrix $\mathbf{A}$.

*Row-based matvec.* The efficiency of row-based matvec is straightforward. For all rows $i = 0, 1, ..., M$:

$$\mathbf{f}'(i) = \sum_{j:\, \mathbf{A}(i,j) \neq 0} \mathbf{A}(i,j) \times \mathbf{f}(j) \quad (2)$$

No matter what the sparsity of $\mathbf{f}$, each row must examine every nonzero, so the number of memory accesses into the matrix required to compute Equation 2 is simply $O(nnz(\mathbf{A}))$.

*Column-based matvec.* However, computing matvec ought to be more efficient if the vector $\mathbf{f}$ is all 0 except for just one element. We define such a situation as *input sparsity*. Can we compute a result without touching all elements in the entire matrix? This is the benefit of column-based matvec: if only $\mathbf{f}(i)$ is nonzero, then $\mathbf{f}'$ is simply the $i$th column of $\mathbf{A}$ i.e., $\mathbf{A}(:, i) \times \mathbf{f}(i)$.

$$\mathbf{f}' = \sum_{i:\, \mathbf{f}(i) \neq 0} \mathbf{A}(:, i) \times \mathbf{f}(i) \quad (3)$$

When $\mathbf{f}$ has more than one non-zero element (when $nnz(\mathbf{f}) > 1$), we must access $nnz(\mathbf{f})$ columns in $A$. How do we combine these multiple columns into the final vector? The necessary operation is a multiway merge of $\mathbf{A}(:, i)\mathbf{f}(i)$ for all $i$ where $\mathbf{f}(i) \neq 0$. Multiway merge (also known as $k$-way merge) is the problem of merging $k$ sorted lists together such that the result is sorted [23]. It arises naturally in column-based matvec from the fact that the outgoing edges of a frontier do not form a set due to different nodes trying to claim the same child. Instead, one obtains $nnz(\mathbf{f})$ lists, and has to solve the problem of merging them together.

Multiway merge takes $n \log k$ memory accesses where $k$ is the number of lists and $n$ is the length of all lists added together. For our problem where we have $k = nnz(\mathbf{f})$ and $n = nnz(m_{\mathbf{f}}^+)$, the multiway merge takes $O(nnz(m_{\mathbf{f}}^+) \log nnz(\mathbf{f}))$.

*Summary.* The complexity of row-based matvec is a constant; we need to touch every element of the matrix even if we want to multiply by a vector that is all 0's except for one index. On the other hand, the complexity of column-based matvec scales with $nnz(m_{\mathbf{f}}^+)$. This matches our intuition, as well as the result of previous work [29], that shows column-based matvec should be more efficient when $\mathbf{f}$ is sparse.

## 3.2 Masked matvec

A useful variant of matvec is *masked matvec*. The intuition behind masked matvec is that it is a mathematical formalism for taking advantage of *output sparsity* (i.e., when we know which elements are zero in the output).

More formally, by masked matvec we mean computing $\mathbf{f}' = (\mathbf{A}\mathbf{f}) .* \mathbf{m}$ where vector $\mathbf{m} \in \mathbb{R}^{M \times 1}$ and $.*$ represents the elementwise multiply operation. This concept of masking gives us a definition for row- and column-based masked matvec. By *row-based*

| Operation | Mask | Expected Cost |
|---|---|---|
| Row- | no | $O(dM)$ |
| based | yes | $O(d\, nnz(\mathbf{m}))$ |
| Column- | no | $O(d\, nnz(\mathbf{f}) \log nnz(\mathbf{f}))$ |
| based | yes | $O(d\, nnz(\mathbf{f}) \log nnz(\mathbf{f}))$ |

Table 1: Four sparse matvec variants and their associated cost, measured in terms of number of memory accesses required in expectation. Note that $d$ is the average number of nonzeroes per row or column.

*masked matvec*, we mean computing for all rows $i = 0, 1, ..., M$:

$$\mathbf{f}'(i) = \begin{cases} \sum_{j:\, \mathbf{A}(i,j) \neq 0} \mathbf{A}(i,j) \times \mathbf{f}(j) & \text{if } \mathbf{m}(i) \neq 0 \\ 0 & \text{if } \mathbf{m}(i) = 0 \end{cases} \quad (4)$$

Similarly for *column-based masked matvec*:

$$\mathbf{f}' = \mathbf{m} .* \sum_{i:\, \mathbf{f}(i) \neq 0} \mathbf{A}(:, i) \times \mathbf{f}(i) \quad (5)$$

The intuition behind masked matvec is that if more elements are masked out (i.e., $\mathbf{m}(i) = 0$ for many indices $i$), then we ought to be doing less work. Looking at the definition above, we no longer need to go through all nonzeroes in $\mathbf{A}$, but merely rows $\mathbf{A}(i, :)$ for which $\mathbf{m}(i) \neq 0$. Thus as shown in Figure 4c, the number of memory accesses decreases to $O(nnz(m_{\mathbf{m}}^-))$. We can avoid $M$ memory accesses by using a data structure similar to the sparse accumulator (SPA) [18] containing both a dense bitmask and a sparse vector containing indices where the zeroes are located. It can be shown that we will only need to form the sparse vector of zeroes once at the cost of $M$ memory accesses, but this cost will be amortized by the number of BFS iterations allowing us to recover $O(nnz(m_{\mathbf{m}}^-))$.

For column-based masked matvec, the number of memory accesses is that of computing column-based matvec, and doing an elementwise multiply with the mask, so the amount of computation does not decrease compared to the unmasked version. At this time, we do not know of an algorithm for column-based matvec that can take advantage of the sparsity of $\mathbf{m}$ and thus reduce the number of memory accesses accordingly.

A summary of the complexity analysis above is shown in Table 1. We choose a matrix ('kron_g500-logn21' from the 10th DIMACS challenge [4]) and perform a microbenchmark to demonstrate the validity of this analysis. We will refer to it as 'kron' henceforth. We use the experimental setup described in Section 7. We measure the runtime of four variants given above for increasing frontier sizes (for the two column-based matvecs), and increasing unvisited node counts (for the two row-based matvecs):

(1) Row-based: increase $nnz(\mathbf{f})$, no mask
(2) Row-based masked: $nnz(\mathbf{f}) = M$, increase $nnz(\mathbf{m})$
(3) Col-based: increase $nnz(\mathbf{f})$, no mask
(4) Col-based masked: increase $nnz(\mathbf{f})$, increase mask at $\frac{2}{3} nnz(\mathbf{f})$

Nodes were selected randomly to belong to the frontier and unvisited nodes. Here, we are using frontier size $nnz(\mathbf{f})$ as a proxy for $nnz(m_{\mathbf{f}})$. The number of outgoing edges $nnz(m_{\mathbf{f}}) \approx d\, nnz(\mathbf{f})$, where $d$ is the average number of outgoing edges per node. Similarly, we use $nnz(\mathbf{m})$ as a proxy for $nnz(m_{\mathbf{m}})$.

The results are shown in Figure 2. They agree with our derivations above. For a given matrix, the row-based matvec's runtime is independent of frontier size and unvisited node count. The runtime of the column-based matvec and the masked row-based matvec both increase with frontier size and unvisited node count, respectively. For low values of either frontier size or unvisited node count, doing either column-based matvec or masked row-based matvec is more efficient than row-based matvec. For high values of either frontier size or unvisited node count, doing the row-based matvec can be more efficient.

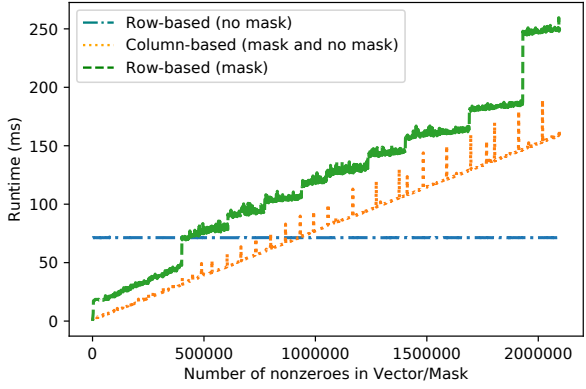

**Figure 2: Runtime in milliseconds for row-based and column-based matvec in their masked and unmasked variants for matrix 'kron' as a function of** $nnz(\mathbf{f})$ **and** $nnz(\mathbf{m})$.

*Structural complement.* Another useful concept is the *structural complement*. Recall the intuition behind masked matvec is that if the mask vector $\mathbf{m}$ is 1 at some index $i$, then it will allow the result of the computation to be passed through to the output $\mathbf{f}'(i)$. The structural complement operator $\neg$ is a user-controlled switch that lets them invert this rule: all the indices $i$ for which $\mathbf{m}$ were 1 will now prevent the result of the computation to be passed through to the output $\mathbf{f}'(i)$, while the indices that were 0 will allow the result ot be passed through.

*Generalized semirings.* One important feature that GraphBLAS provides is that it allows users to express different traversal graph algorithms such as BFS, SSSP (Bellman-Ford), PageRank, maximal independent set, etc. using matvec and matmul [21]. This way, the user can succinctly express the desired graph algorithm in a way that makes parallelization easy. This is analogous to the key role Level 3 BLAS (Basic Linear Algebra Subroutines) plays in scientific computing; it is much easier to optimize for a set of standard operations than have scientists optimize every application all the way down to the hardware-level. The mechanism in which they are able to do so is called *generalized semirings*.

What generalized semirings do is allow the user to replace the standard matrix multiplication and addition operation over the real number field with zero-element 0 ($\mathbb{R}, \times, +, 0$) by any operation they want over arbitrary field $\mathbb{D}$ with zero-element $\mathbb{I}$ ($\mathbb{D}, \otimes, \oplus, \mathbb{I}$). We refer to the latter as *matvec over semiring* ($\mathbb{D}, \otimes, \oplus, \mathbb{I}$). We also have the row-based and column-based equivalents for all semirings. For example, *row-based matvec over semiring* ($\mathbb{D}, \otimes, \oplus, \mathbb{I}$) is:

$$\mathbf{f}'(i) = \bigoplus_{\substack{\mathbf{A}(i,j) \neq \mathbb{I} \\ j=0}}^{n} \mathbf{A}(i,j) \otimes f(j)$$

For row-based masked and column-based masked matvec over semirings, we generalize the element-wise operation to be $\odot: \mathbb{D} \times \mathbb{D}_2 \to \mathbb{D}$ where $\mathbb{D}_2$ is the set of allowable values of the mask vector $\mathbf{m}$ and $\mathbb{D}$ is the set of allowable values of the matrix $\mathbf{A}$ and vector $\mathbf{f}$. For example, *row-based masked matvec over semiring* ($\mathbb{D}, \otimes, \oplus, \mathbb{I}$) and element-wise multiply $\odot: \mathbb{D} \times \mathbb{D}_2 \to \mathbb{D}$ is:

$$\mathbf{f}'(i) = \mathbf{m}. \odot \bigoplus_{\substack{\mathbf{A}(i,j) \neq \mathbb{I} \\ j=0}}^{n} \mathbf{A}(i,j) \otimes f(j)$$

## 4 RELATING MATVEC AND PUSH-PULL

In this section, we discuss the connection between masked matvec and the three optimizations inherent to DOBFS. Then, we discuss two closely related optimizations that were not in the initial direction-optimization paper [7] or its successor [8], which looked at matvec in the context of row- and column-based variants for PageRank. This later work examines three blocking methods (cache, propagation, and deterministic propagation) for computing matvec using row- and column-based approaches. Besta et al. also observed the duality between push-pull and row- and column-based matvec in the context of several graph algorithms. They give a theoretical analysis on three parallel random access memory (PRAM) variants for differences between push-pull. We extend their push-pull analysis to include the concept of masking, which is needed to take advantage of output sparsity and express early exit.

### 4.1 Connection with push

To demonstrate the connection with push, we consider the formulation of the problem using $\mathbf{f}' = \mathbf{A}^\mathbf{T}\mathbf{f} .* \neg\mathbf{v}$ in the specific context of one BFS iteration. In graphical terms, this is visualized as Figure 3a. Our current frontier is shown by nodes marked in orange. The visited vector $\mathbf{v}$ indicates the already visited nodes $A, B, C, D$.

We will first consider the push case as shown in Figure 3d. We must examine all edges leaving the current frontier. Doing so, we examine the children of $B, C, D$, and combine them using a logical OR. This allows us to discover nodes $A, E, F$. From these 3 discovered nodes, we eliminate the already-discovered node $A$ from our frontier by filtering using the visited vector $\mathbf{v}$. This leaves us with the two nodes marked in dark purple $E, F$ as the newly discovered nodes. In matvec terms, our operation is the same: we find the neighbors of the current frontier (represented by columns of $\mathbf{A}^\mathbf{T}$) and merge them together before filtering using $\mathbf{v}$. This is a well-known result [11, 17].

### 4.2 Connection with pull

Now let us consider the pull case (shown in Figure 3c). Here, the traversal pattern is different, because we must take the *unvisited* vertices $\neg\mathbf{v}$ as our starting point (Optimization 2: masking). We start from each unvisited vertex $E, F, G, H$ and look at each node's parents. For any unvisited vertex, once a single parent has been found to belong in the visited vector, we can mark the node as discovered and stop visiting its parents (Optimization 3: early-exit). In matvec terms, we apply the unvisited vector $\mathbf{v}$ as a mask to our matrix to take advantage of *output sparsity*. Since we know that the first four nodes with values (0, 1, 1, 1) will be filtered out, we can skip computing matvec for them. For the rest, we will begin

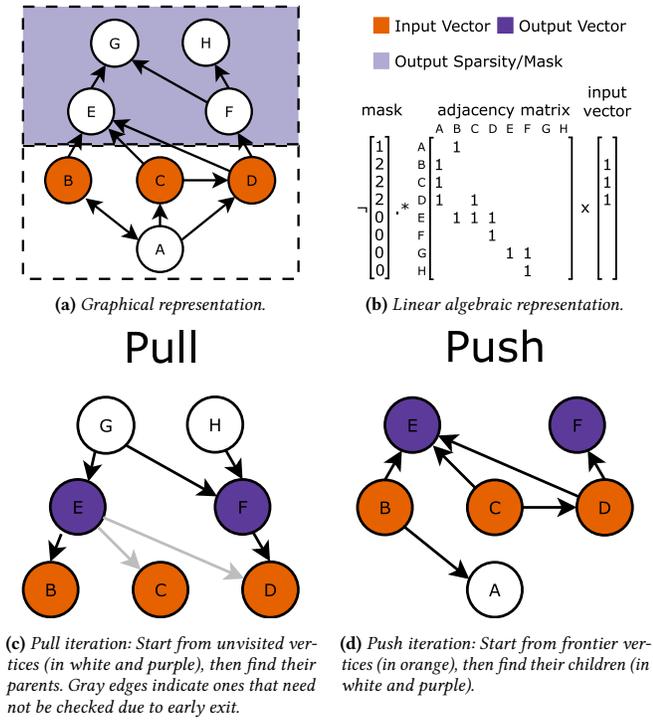

(a) *Graphical representation.*  (b) *Linear algebraic representation.*

## Pull

(c) *Pull iteration: Start from unvisited vertices (in white and purple), then find their parents. Gray edges indicate ones that need not be checked due to early exit.*

## Push

(d) *Push iteration: Start from frontier vertices (in orange), then find their children (in white and purple).*

**Figure 3: Simple example showing BFS traversal from the 3 nodes marked orange. There is a one-to-one correspondence between the graphical representation of both traversal strategies and their respective matvec equivalents in Fig. 4.**

examining each unvisited node's parents until we find one that is in the frontier. Once we have found one, it is possible to early-exit.

In mathematical terms, performing the early-exit is justified inside an matvec inner loop as long as the addition operation of the matvec semiring is an OR that evaluates to true. This is the same principle by which C compilers allow short-circuit evaluation. This can easily be implemented in the GraphBLAS implementation by checking whether the matvec semiring is logical OR.

Given the treatment above, we observe that the pull case can be expressed by the same formula $\mathbf{f}' = \mathbf{A}^T \mathbf{f} .* \neg \mathbf{v}$ as the push case that is already part of the GraphBLAS C API [13]. The full BFS algorithm is shown in Algorithm 1. Our observation that both push and pull use the same formulation allows the developer to express DOBFS's push and pull using that single formulation; the GraphBLAS backend can then make a runtime decision whether to use the column-based matvec or the row-based matvec (Optimization 1: change of direction) to achieve maximum efficiency.

## 5 OPTIMIZATIONS

In this section, we discuss in depth the five optimizations mentioned in the previous section. We also analyze their suitability for generalization to speeding up matvec for other applications.

(1) Change of direction
(2) Masking
(3) Early-exit

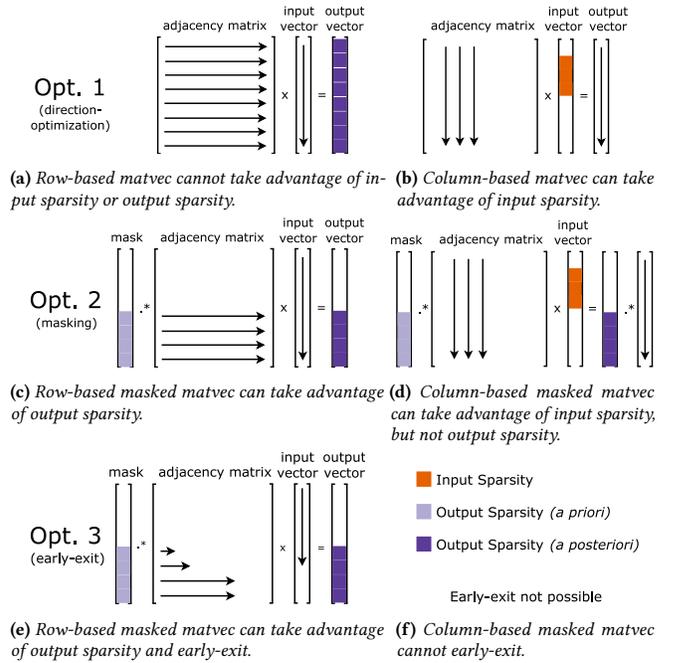

(a) *Row-based matvec cannot take advantage of input sparsity or output sparsity.*  (b) *Column-based matvec can take advantage of input sparsity.*

(c) *Row-based masked matvec can take advantage of output sparsity.*  (d) *Column-based masked matvec can take advantage of input sparsity, but not output sparsity.*

(e) *Row-based masked matvec can take advantage of output sparsity and early-exit.*  (f) *Column-based masked matvec cannot early-exit.*

**Figure 4: The three optimizations known as "direction-optimized" BFS. We are the first to generalize Optimization 2 by showing that masking can achieve asymptotic speed-up over standard row-based matvec when output sparsity is known before computation (i.e., *a priori*).**

---

**Algorithm 1** BFS using Boolean semiring ({0, 1}, OR, AND, 0) with equivalent GraphBLAS operations highlighted in comments. Shown pseudocode implements both push and pull, so it is up to Optimization 1: change of direction in the backend to decide which one is more efficient to use. For GrB_mxv, the operations are changed from their standard matrix multiplication meaning to become × = AND, + = OR. GrB_assign and GrB_reduce uses the standard matrix multiplication meanings for the × and +.

1: **procedure** GRB_BFS(Vector **v**, Graph **A**, Source *s*)
2:　　Initialize $d \leftarrow 1$
3:　　Initialize $\mathbf{f}(i) \leftarrow \begin{cases} 1, & \text{if } i = s \\ 0, & \text{if } i \neq s \end{cases}$　　▷ GrB_Vector_new
4:　　Initialize $\mathbf{v} \leftarrow [0, 0, \ldots, 0]$　　▷ GrB_Vector_new
5:　　Initialize $c \leftarrow 1$
6:　　**while** $c > 0$ **do**
7:　　　　Update $\mathbf{v} \leftarrow \mathbf{f} \times d + \mathbf{v}$　　▷ GrB_assign
8:　　　　Update $\mathbf{f} \leftarrow \mathbf{A}^T \mathbf{f} .* \neg \mathbf{v}$　　▷ GrB_mxv
9:　　　　Compute $c \leftarrow \sum_{i=0}^{n} \mathbf{f}(i)$　　▷ GrB_reduce
10:　　　Update $d \leftarrow d + 1$
11:　　**end while**
12: **end procedure**

---

(4) Operand reuse
(5) Structure-only

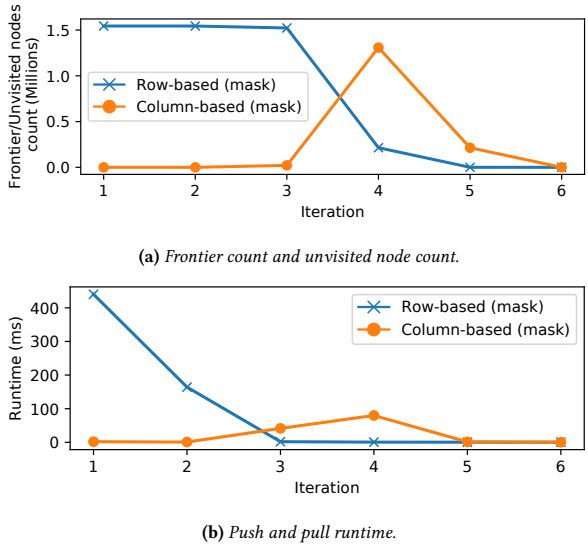

(a) *Frontier count and unvisited node count.*

(b) *Push and pull runtime.*

**Figure 5: Breakdown of edge types in frontier during BFS traversal of Kronecker scale-21 graph (2M vertices, 182M edges).**

Optimizations 1, 2 and 3 form the commonly used definition of DOBFS from the paper that first described it [7]. Optimizations 4 and 5 also contribute to a high-performance BFS. The impact of these five optimizations are summarized in Table 2.

### 5.1 Optimization 1: Change of direction

When the frontier becomes large, instead of each frontier node looking for its children and adding them to the next frontier (*push*), it becomes more efficient if each *unvisited* node looks for its parents (*pull*). Near the end of the computation, the number of frontier edges once again falls, and it is profitable to return once more to *push*. Efficient DOBFS traversals on scale-free graphs result in three distinct phases:

(1) Push phase: Frontier is small, unvisited vertices is large.
(2) Pull phase: Frontier is medium, unvisited vertices is large.
(3) Push phase: Frontier is small, unvisited vertices is small.

Figure 5 shows an empirical representation of this phenomenon on a Kronecker graph of scale-21 with 2M vertices and 182M edges. In Iterations 1–2 of the BFS, the frontier size is small. Similarly, the number of unvisited vertices is big, so it is profitable to use *push*. In Iteration 6, the frontier size falls once more, so it is worthwhile to go back to *push*. The frontier size and number of unvisited vertices are comparable for Iterations 2 and 3. However, the performance of row-based-with-mask and column-based-with-mask is drastically different between these two iterations. Row-based-with-mask runtime drops precipitously, but column-based-with-mask runtime increases.

To solve this problem, we perform another microbenchmark that differs in two respects compared to Figure 2. First, in the previous benchmark, we generated random vectors as input and mask; here we launch BFS from 1000 different sources and plot the per-iteration runtime as a function of input frontier size (in the case of column-based) and unvisited nodes (in the case of row-based-with-mask), so the vectors have semantic meaning. Second, for row-based-with-mask we activate the early exit optimization. The result of the microbenchmark is shown in Figure 6. Interestingly, the runtimes of the row-based and column-based matvecs look very different depending on whether the input vector is random or whether it represents a BFS frontier.

We begin by analyzing the column-based-with-mask case (the red oval in Figure 6). This interesting shape is characteristic of power-law graphs (of which 'kron' is a member) that have a small number of nodes with many edges (which we term supervertices) but most nodes only a few. We examined a few examples of the runtime data. Column-based BFS progresses with increasing iteration count in a clockwise direction. In the first phase of the BFS, the algorithm quickly discovers a few supervertices, which dramatically increases its runtime. The BFS then reaches a peak in the frontier size (Iteration 4 of Figure 5a), at which point the frontier begins to fall. Its return to low frontier size corresponds to the bottom of the oval. Despite a comparable nonzero count in the input vector, this phase is notable for its lack of supervertices, which keeps runtime to a minimum.

Row-based-with-mask has a different pattern. There, the row-based BFS begins at the top of the backwards 'L', then moves down and towards the left with increasing iteration count. When there have only been one or two nodes visited, we most likely have not discovered a supervertex yet, so the runtime is high (above 100 ms in Figure 6). We found it took on average 79 parent checks before a valid parent was found. After the first 20k nodes have been visited, it is with high probability that a supervertex has been visited. In either case, we found that after 2 BFS iterations, the average number of parents examined before a valid parent was found dropped from 79 to 1.3. This supervertex concept does not exist in the matvec of Figure 2, so there we get a different result. The line peaks around 1.5M, because that is the size of the largest connected component.

In light of DOBFS, it becomes clear from looking at Figure 6b that at the start of the BFS, the row-based BFS is below the blue line around $(0, 10^{-1})$, but the row-based BFS is above the blue line, so it is more efficient to do row-based BFS for the first few iterations. By Iteration 3, the row-based has increased to near the blue line, while the row-based has dropped sharply below. At this point, it is more efficient to do column-based BFS. Near the end of the algorithm, both algorithms continue to improve in efficiency, so either algorithm will suffice.

### 5.2 Optimization 2: Masking

As described in Section 3, masking means computing only the rows whose value we know *a priori* must be updated. In other words, depending on the *output sparsity* as expressed by the mask, we can perform a matvec using a subset of the memory accesses required by the unmasked variant. This yields the algorithmic speed-up of $O(d\ nnz(\mathbf{m}))$ (masked) compared to $O(dM)$ (unmasked). As an example, Figure 4c shows that instead of computing row-based matvec for all 8 rows, we can reduce computation to only the bottom 4 rows and obtain the same result.

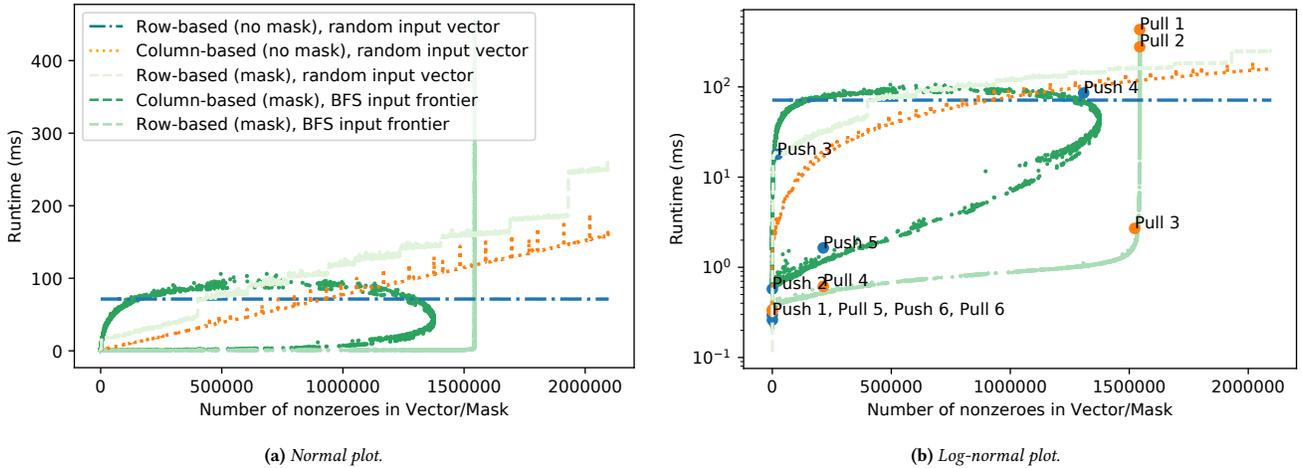

Figure 6: Runtime in milliseconds for row-based and column-based matvec in their masked and unmasked variants for matrix 'kron' as a function of *nnz*(f) and *nnz*(m). Input vectors and masks are generated using 2 methods: (1) random input vectors, (2) based on sampling BFS iterations from 1000 random sources on graph 'kron_g500-logn21'. Push 1 means Iteration 1 of the push-only BFS. Pull 1 means Iteration 1 of the pull-only BFS. For this graph, 2 iterations of push followed by 3 of pull, then 1 iteration of push or pull, yields the best performance.

### 5.3 Optimization 3: Early-exit

In the pull phase, an undiscovered node searches for a parent in the frontier. Once such a parent has been found, further computation for that undiscovered node is unnecessary and can potentially halt. For the nodes who do not have parents that have been previously visited, early-exit has no benefit. Our results (Table 2) indicate this optimization yielded the greatest speed-up.

### 5.4 Optimization 4: Operand reuse

Since the set of the visited node list is always a superset of the frontier node list, we can simply use the visited node list in place of the frontier. Gunrock [31] notes that $f \subset v$ and computes $A^T v .* \neg v$ instead of $A^T f .* \neg v$. This is a helpful optimization, because computing the latter means that during the iteration in which we are switching from push to pull, we get the costly sparse-to-dense frontier vector conversion for free because in the above expression, the frontier $f$ is not required as part of the input.

### 5.5 Optimization 5: Structure-only

The matrix values and sparse vector values do not need to be accessed for BFS. This optimization takes advantage of the fact for the purposes of a BFS, the matrix can be implicitly treated as a Boolean matrix, because we treat the existence of sparse matrix column indices as a Boolean 1, and non-existence as Boolean 0. This optimization only applies to the push phase, and the majority of the speed-up comes during the multiway merge. In Section 6, we say that we implement this multiway merge using a radix sort. This radix sort is often the bottleneck of the algorithm, so if we use this optimization, we are reducing a key-pair sort to a key-only sort, which will reduce the number of memory accesses by a factor of 2.

| Optimization | Performance (GTEPS) | Speed-up |
|---|---|---|
| Baseline | 0.874 | — |
| Structure only | 1.411 | 1.62× |
| Change of direction | 1.527 | 1.08× |
| Masking | 3.932 | 2.58× |
| Early exit | 15.83 | 4.02× |
| Operand reuse | 42.44 | 2.68× |

Table 2: Impact of the four optimizations described in this section on the performance measured in billions of traversed edges per second on 'kron_g500-logn21'. These optimizations are cumulative, meaning the next optimization is stacked on top of the previous one. Speedups are standalone.

### 5.6 Generality

Change of direction can be generalized to other algorithms, including betweenness centrality, personalized PageRank, and SSSP, with similar tradeoffs between row-based and column-based approaches (Table 1). In SSSP, for example, despite the workfront evolution (how the frontier changes between iterations) being completely different from Figure 5a, a simple 2-phase direction-optimized traversal can be used where the traversal is begun using unmasked column-based matvec, with a switch to row-based matvec when the frontier becomes large enough that row-based is more efficient. We describe how to implement change of direction in Section 6.3.

Masking can be generalized to any algorithm where the *output sparsity* is known before the operation. This includes algorithms such as triangle counting and enumeration, adaptive PageRank, batched betweenness centrality, maximal independent set, and sparse neural networks. In all of these algorithms, an optimal algorithm will want to use the knowledge that some output elements will be zero (e.g., when the PageRank value has converged for a particular node). In these cases, our proposed elementwise multiply formalism provides the mathematical theory to take advantage of this *output sparsity*, yielding an asymptotic speed-up of $O(\frac{dM}{d\,nnz(\mathbf{m})}) = O(\frac{M}{nnz(\mathbf{m})})$.

Operand reuse is generalizable to any traversal-based algorithm for which computing $\mathbf{A}^T\mathbf{v}$ in place of $\mathbf{A}^T\mathbf{f}$ gives the correct result. We give SSSP and personalized PageRank as examples for which this holds true. However, early-exit and structure-only are only generalizable to semirings that operate on Booleans.

## 6 IMPLEMENTATION

This section will discuss our GPU-based implementation of row-based masked matvec, column-based matvec (masked and unmasked), and our direction-optimization heuristic. For simplicity in the following discussion, we use $\mathbf{m}(i)$ to denote checking whether the mask is nonzero, and if so, allowing the value to pass through to the output if it is. $\neg\mathbf{m}(i)$, while not discussed, does the inverse.

### 6.1 Row-based masked matvec (Pull phase)

Our parallel row-based masked matvec on the GPU is listed in Algorithm 2 and illustrated in Figure 4e. We parallelize over threads and have each thread check the mask $\mathbf{m}$. If $\mathbf{m}(i)$ passes the check, the thread $i$ checks its neighbors $j$ in the matrix $\mathbf{A}^T(i,:)$ and tallies up the result if and only if the $\mathbf{v}(j)$ is also nonzero. For semirings with Boolean operators that support short-circuiting such as the one used for BFS, namely the Boolean semi-ring ($\{0, 1\}$, AND, OR, 0), once a single non-zero neighbor $j$ is discovered (meaning it has been visited before), the thread can immediately write its result to the output vector and exit.

**Algorithm 2** Masked row-based matrix multiplication over the generalized semiring ($\mathbb{D}, \otimes, \oplus, \mathbb{I}$). The Boolean variable scmp controls whether or not $\mathbf{m}$ or $\neg\mathbf{m}$ is used.

1: **procedure** ROW_MASKED_MXV(Vector $\mathbf{v}$, Graph $\mathbf{A}^T$, MaskVector $\mathbf{m}$, MaskIdentity identity, Boolean scmp, Boolean accum)
2:   **for** each thread $i$ in parallel **do**
3:     **if** $\mathbf{m}(i) \ne$ identity XOR scmp **then**
4:       value $\leftarrow \mathbb{I}$
5:       **for** index $j$ in $\mathbf{A}^T(i,:)$ **do**
6:         **if** $\mathbf{v}(j) \ne 0$ **then**
7:           value $\leftarrow$ value $\oplus \mathbf{A}^T(i,j) \otimes \mathbf{v}(j)$
8:           **break** (optional: *early-exit* opt. enables this break)
9:         **end if**
10:       **end for**
11:       $\mathbf{w}(i) \leftarrow$ accum ? $\mathbf{w}(i)+$ value : value
12:     **end if**
13:   **end for**
14: **return** $\mathbf{w}$
15: **end procedure**

### 6.2 Column-based masked matvec (Push phase)

Our column-based masked matvec follows Gustavson's algorithm for SpGEMM (sparse matrix-sparse matrix multiplication), but specialized to matvec [19]. The key challenge in parallelizing Gustavson's algorithm is solving the multiway merge problem [1]. For the GPU, our parallelization approach follows the scan-gather-sort approach outlined by Yang et al. [32] and is shown in Algorithm 3. Instead of doing the multiway merge by doing $O(nnz(m_{\mathbf{f}}^+) \log nnz(\mathbf{f}))$, we concatenate all lists and use radix sort, because radix sort tends to have better performance on GPUs. Our complexity then becomes $O(nnz(m_{\mathbf{f}}^+) \log M)$, where M is the number of rows in the matrix; an increase in M forces us to do a higher-bit radix sort.

We begin by computing the requisite space to leave in the output frontier for each neighbor list expansion. In compressed sparse row (CSR) format, node $i$ computes its required space by taking the difference between the $i$-th and $i+1$-th row pointer values. Once each thread has its requisite length, we perform a prefix-sum over these lengths. This is fed into a higher-level abstraction, INTERVALGATHER, from the ModernGPU library [5]. On the prefix-sum array, INTERVALGATHER does a parallel search on sorted input to determine the indices from which each thread must gather. This gives us a load-balanced way of reading the column indices and values (Lines 6–9 in Algorithm 3).

During this process, the vector value of the corresponding thread $\mathbf{v}(i)$ is also gathered from global memory. This will allow us to multiply all of $i$'s neighbors with $\mathbf{v}(i)$ using the $\otimes$ operator. Once this is done, we write the column indices and multiplied values to global memory. Then we run a log $M$-bit radix sort, where $M$ is the number of matrix rows (as mentioned in Section 2). One advantage of the structure-only optimization is that it allows us to cut down on the runtime, because this radix sort is often the bottleneck of the column-based masked matvec.

After the radix sort, a segmented reduction using the operator ($\oplus, \mathbb{I}$) gives us the temporary vector. The unmasked column-based matvec ends here. The masked version additionally filters out the values not in the mask by checking $\mathbf{m}(i)$.

### 6.3 Direction-optimization heuristic

Implementing an efficient DOBFS requires good decisions to switch between forward and reverse. Beamer et al. proposed a heuristic to switch from push to pull when $\frac{nnz(m_{\mathbf{f}})}{nnz(m_{\mathbf{u}})} > \alpha$ for some factor $\alpha$, and to switch back when $\frac{nnz(\mathbf{f})}{M} < \beta$ for some factor $\beta$ [7]. We aim to match their intent but also wish to avoid computing $m_f$ speculatively. Instead, we rely on the fact that $nnz(m_f) \approx d\,nnz(\mathbf{f})$, where $d$ is the average number of nonzeroes per row of the matrix and $M$ is the number of rows in the matrix. If we also assume that $nnz(m_{\mathbf{u}}) \approx nnz(\mathbf{A}) \approx dM$ when we desire to switch, we see that $\frac{nnz(m_{\mathbf{f}})}{nnz(m_{\mathbf{u}})} \approx \frac{d\,nnz(\mathbf{f})}{dM} = \frac{nnz(\mathbf{f})}{M} = r$. Our method thus reduces to $\alpha = \beta$; if $r$ is increasing and $r > \alpha$, we switch from push to pull, and if $r$ is decreasing and $r < \beta$, then we switch from pull to push. In this paper, we use $\alpha = \beta = 0.01$, which is optimal (in comparison to computing both push and pull on each iteration and choosing the fastest) for graphs we studied except 'i04' and the 3 non-scale free graphs, whose optimal BFS is push-only for all iterations.

To decide which version of matvec to use, we call the CONVERT function on the input vector $\mathbf{f}$. Then the vector tests whether $\mathbf{f}$ is stored in DenseVector or SparseVector format. If the former, then it checks whether the number of nonzeroes is low enough to warrant converting to a SparseVector using DENSE2SPARSE and whether it has decreased from the last time CONVERT was called on it. If the latter, it checks whether the number of nonzeroes is high enough

**Algorithm 3** Masked column-based matrix multiplication over generalized semiring $(\mathbb{D}, \otimes, \oplus, \mathbb{I})$. The Boolean variable scmp controls whether or not **m** or ¬**m** is used. The Boolean variable accum controls whether accumulat

1: **procedure** COL_MASKED_MXV(Vector **v**, Graph $A^T$, MaskVector **m**, MaskIdentity identity, Boolean scmp, Boolean accum)
2:   **for** each thread $i$ **in parallel do**
3:     length($i$) ← row_ptr($i$+1)-row_ptr($i$) for all i such that **v**($i$) ≠ $\mathbb{I}$
4:   **end for**
5:   scan ← prefix-sum length
6:   addr($i$) ← INTERVALGATHER(scan, **v**)
7:   col ← col_ind($j$) such that $A^T(j, i) \neq \mathbb{I}$ from addr($i$)
8:   val ← $A^T(j, i)$ from addr($i$)
9:   val ← val ⊗ **v**($i$)
10:   write (col, val) to global memory
11:   Barrier synchronization
12:   key-value sort (col, val)
13:   (optional: *structure only* opt. turns this into a key-only sort)
14:   Barrier synchronization
15:   segmented-reduction using $(\oplus, \mathbb{I})$ produces **w**′
16:   Barrier synchronization
17:   **for** each thread $i$ **in parallel do**
18:     ind ← ind such that **w**′(ind) ≠ $\mathbb{I}$
19:     **if** **m**(ind) ≠ identity XOR scmp **then**
20:       **w**″($i$) = **w**′($i$)
21:     **else**
22:       **w**″($i$) = $\mathbb{I}$
23:     **end if**
24:   **end for**
25:   **if** accum **then**
26:     **w** ← **w** + **w**″
27:   **else**
28:     **w** ← **w**″
29:   **end if**
30:   **return w**
31: **end procedure**

to warrant converting to a DenseVector using SPARSE2DENSE and whether it has increased since the last time CONVERT was called. The user can select this sparse/dense switching point by passing in a floating-point value through the Descriptor of the matvec call. The default switchpoint is when the ratio of nonzeroes in the sparse matrix exceeds 0.01. Another way of expressing this is that once we have visited 1% of vertices in the graph in a BFS, we are sure to have hit a supernode.

As mentioned in Section 3, it is more efficient to store the input vector as a DenseVector object for row-based matvec. Similarly, it is efficient to store the frontier vector as a SparseVector. Therefore switching from push-to-pull in our implementation means converting the input vector from sparse to dense, and vice versa for pull-to-push. Using these SparseVector and DenseVector objects, we have the function signatures of following operations, which correspond to the four variants analyzed in Table 1:

(1) ROW_MXV( DenseVector w, GrB_NULL, Matrix A, DenseVector v )
(2) ROW_MASKED_MXV( DenseVector w, DenseVector mask, Matrix A, DenseVector v )
(3) COL_MXV( SparseVector w, GrB_NULL, Matrix A, SparseVector v )
(4) COL_MASKED_MXV( SparseVector w, DenseVector mask, Matrix A, SparseVector v )

| Dataset | Vertices | Edges | Max Degree | Diameter | Type |
|---|---|---|---|---|---|
| soc-orkut | 3M | 212.7M | 27,466 | 9 | rs |
| soc-Livejournal1 | 4.8M | 85.7M | 20,333 | 16 | rs |
| hollywood-09 | 1.1M | 112.8M | 11,467 | 11 | rs |
| indochina-04 | 7.4M | 302M | 256,425 | 26 | rs |
| kron_g500-logn21 | 2.1M | 182.1M | 213,904 | 6 | gs |
| rmat_s22_e64 | 4.2M | 483M | 421,607 | 5 | gs |
| rmat_s23_e32 | 8.4M | 505.6M | 440,396 | 6 | gs |
| rmat_s24_e16 | 16.8M | 519.7M | 432,152 | 6 | gs |
| rgg_n_24 | 16.8M | 265.1M | 40 | 2622 | gm |
| roadNet_CA | 2M | 5.5M | 12 | 849 | rm |
| road_USA | 23.9M | 577.1M | 9 | 6809 | rm |

Table 3: Dataset Description Table. Graph types are: r: real-world, g: generated, s: scale-free, and m: mesh-like.

There are already many efficient implementations of Operation (1) on the GPU [5, 9, 26]—we use ModernGPU's SpmvCsrBinary—but we implemented the other 3 operations ourselves.

## 7 EXPERIMENTAL RESULTS
### 7.1 Experimental setup

We ran all experiments[1] in this paper on a Linux workstation with 2× 3.50 GHz Intel 4-core E5-2637 v2 Xeon CPUs, 556 GB of main memory, and an NVIDIA K40c GPU with 12 GB on-board memory. The GPU programs were compiled with NVIDIA's nvcc compiler (version 8.0.61). The C code was compiled using gcc 4.9.4. Ligra was compiled using icpc 15.0.1 with CilkPlus. All results ignore transfer time (from disk-to-memory and CPU-to-GPU). The gather, scan, and row-based mxv (without mask) operations are from the ModernGPU library [5]. The radix sort and reduce operations are from the CUDA UnBound library (CUB) [25]. All BFS tests were run 10 times with the average runtime and MTEPS used for results.

The datasets we used are listed in Table 3. soc-ork is from Network Repository [28]; soc-lj, h09, i04, kron, roadNet_CA, and road_usa are from the UF Sparse Matrix Collection [16]; and rmat-22, rmat-23, rmat-24, and rgg are randomized graphs we generated. All datasets have been converted to undirected graphs. Self-loops and duplicated edges are removed.

### 7.2 Graph framework comparison

As a baseline for comparison, we use the push-based BFS on the GPU by Yang et al. [32], because it is based in linear algebra and is (relatively) free of graph-specific optimizations. It does not support DOBFS. We also compare against four other graph frameworks (1 linear-algebra-based, 3 native-graph). SuiteSparse is a single-threaded CPU implementation of GraphBLAS. It is notable for being the first GraphBLAS implementation that adheres closely to the specification [15]. SuiteSparse performs matvecs with the column-based algorithm. CuSha is a vertex-centric framework on the GPU using the gather-apply-scatter (GAS) programming model [22]. Ligra is a CPU-based vertex-centric framework for shared memory [29]. It is the fastest graph framework we found on a multi-threaded CPU and was the first work that generalized push-pull to traversal algorithms other than BFS. Gunrock is a GPU-based

---
[1]https://github.com/owensgroup/push-pull

| Dataset | Runtime (ms) [lower is better] | | | | | | Edge throughput (MTEPS) [higher is better] | | | | | |
|---|---|---|---|---|---|---|---|---|---|---|---|---|
| | SuiteSparse | CuSha | Baseline | Ligra | Gunrock | This Work | SuiteSparse | CuSha | Baseline | Ligra | Gunrock | This Work |
| soc-ork | 2165 | 244.9 | 122.4 | 26.1 | **5.573** | 7.280 | 98.24 | 868.3 | 1722 | 8149 | **38165** | 29217 |
| soc-lj | 1483 | 263.6 | 51.32 | 42.4 | **14.05** | 14.16 | 57.76 | 519.5 | 1669 | 2021 | **6097** | 6049 |
| h09 | 596.7 | 855.2 | 23.39 | 12.8 | **5.835** | 7.138 | 188.7 | 131.8 | 4814 | 8798 | **19299** | 15775 |
| i04 | 1866 | 17609 | **71.81** | 157 | 77.21 | 80.37 | 159.8 | 22.45 | **4151** | 1899 | 3861 | 3709 |
| kron | 1694 | 237.9 | 108.7 | 18.5 | 4.546 | **4.088** | 107.5 | 765.5 | 1675 | 9844 | 40061 | **44550** |
| rmat-22 | 4226 | 1354 | OOM | 22.6 | **3.943** | 4.781 | 114.3 | 369.1 | OOM | 21374 | **122516** | 101038 |
| rmat-23 | 6033 | 1423 | OOM | 45.6 | **7.997** | 8.655 | 83.81 | 362.7 | OOM | 11089 | **63227** | 58417 |
| rmat-24 | 8193 | 1234 | OOM | 89.6 | 16.74 | **16.59** | 63.42 | 426.4 | OOM | 5800 | 31042 | **31327** |
| rgg | 230602 | 68202 | 9147 | 918 | **593.9** | 2991 | 1.201 | 3.887 | 30.28 | 288.8 | **466.4** | 92.59 |
| roadnet | 342 | 288.5 | 284.9 | **82.1** | 130.9 | 214.4 | 16.14 | 14.99 | 19.38 | **67.25** | 42.18 | 25.75 |
| road_usa | 9413 | 36194 | 26594 | 978 | **676.2** | 7155 | 6.131 | 7.944 | 2.17 | 59.01 | **85.34** | 8.065 |

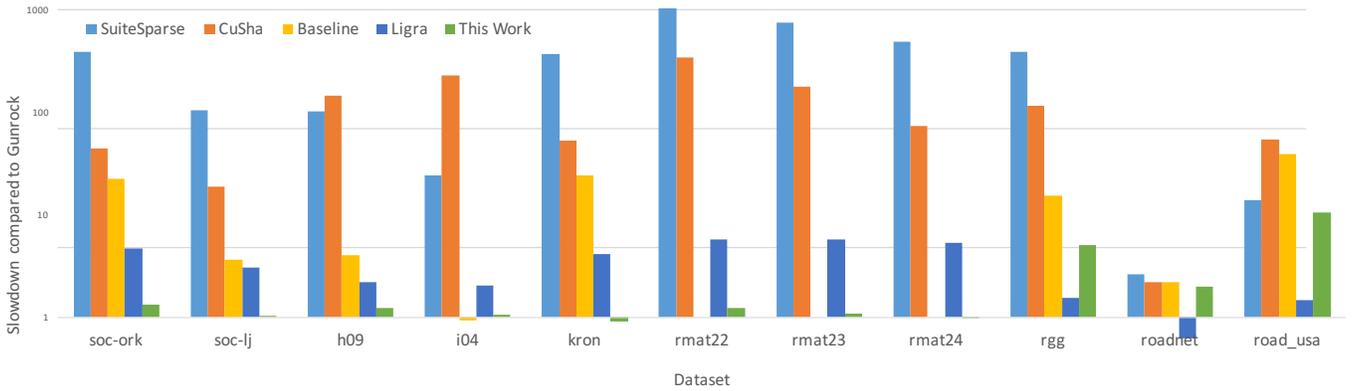

Figure 7: Comparison of our work to other graph libraries (SuiteSparse [15], CuSha [22], a baseline push-based BFS [32], Ligra [29], and Gunrock [31]) implemented on 1× Intel Xeon 4-core E5-2637 v2 CPU and 1× NVIDIA Tesla K40c GPU. Bold is fastest for that dataset. OOM means out of memory. The graph presents the same data as a slowdown compared to Gunrock.

frontier-centric framework [31] that generated the fastest single-processor BFS in our experiments.

### 7.3 Discussion of results

Figure 7 shows our performance results. In terms of runtime, we are 122×, 48.3×, 3.37×, 1.16× faster in the geomean than SuiteSparse, CuSha, the baseline, and Ligra respectively. We are 34.6% slower than Gunrock. Our implementation is relatively better on scale-free graphs, where we are 3.51× faster than Ligra on the scale-free datasets. In comparison, we are 3.2× slower than Ligra on the road maps and mesh graph. Our performance with respect to Gunrock is similar in that we do poorly on road maps (3.15× slower) compared with scale-free graphs (1.09× slower). This supports our intuition from Section 4 that DOBFS is helpful mainly on scale-free graphs.

The four biggest differences between Gunrock's and our implementation is that on top of the optimizations discussed in this paper, they also employ (1) local culling, (2) keeping sparse vector indices in unsorted order with possible duplicates, (3) kernel fusion, and (4) a different traversal strategy for road networks.

*Local culling.* Instead of our write to global memory in Line 9 of Algorithm 3 which is followed by an expensive key-value sort, Gunrock's filter step incorporates a series of inexpensive heuristics [27] to reduce but not eliminate redundant entries in the output frontier. These heuristics include a global bitmask, a block-level history hashtable, and a warp-level hashtable. The size of each hashtable is adjustable to achieve the optimal tradeoff between performance and redundancy reduction rate. However, this approach may not be suitable for GraphBLAS, because such an optimization may be too BFS-focused and would generalize poorly.

*Unsorted order and redundant elements.* When performing the column-based masked matvec as in Figure 4d, our complexity is $O(nnz(m_f^+) \log M)$, so the bottleneck is in making the elements unique. If duplicate indices are tolerated, we can omit the multiway merge entirely, and get rid of the logarithmic factor leaving us with $O(nnz(m_f^+))$. While redundant vertices impose an extra cost, BFS is an algorithm that can tolerate redundant vertices and in some cases, it may be cheaper to allow a few extra vertices to percolate through the computation than to go to significant effort to filter them out. This approach may not be suitable for GraphBLAS, because such an optimization may be too BFS-focused and would generalize poorly.

*Kernel fusion.* Because launching a GPU kernel is relatively expensive, optimized GPU programs attempt to fuse multiple kernels into one to improve performance ("kernel fusion"). Gunrock fuses kernels in several places, for example, fusing Lines 7 and 8 in Algorithm 1 during pull traversal. This optimization may be a good fit for a non-blocking implementation of GraphBLAS, which would construct a task graph and fuse tasks to improve performance.

*Different traversal strategy for road networks.* For road networks, Gunrock uses the TWC (Thread Warp CTA) load-balancing mechanism of Merrill et al. [27]. TWC is cheaper to apply than other load-balancing mechanisms, which makes it a good match for road networks that have many BFS iterations each with little work.

## 8 CONCLUSION

In this paper we demonstrate that push-pull corresponds to the concept of column- and row-based masked matvec. We analyze four variants of matvec, and show theoretically and empirically they have fundamentally different computational complexities. We provide experimental evidence that the concept of a mask to take advantage of *output sparsity* is critical for a linear-algebra based graph framework to be competitive with state-of-the art vertex-centric graph frameworks on parallel GPU and CPUs.

A possible future research direction would be to use this work as a building block for a distributed GraphBLAS implementation on GPUs. Extending this work to other graph algorithms such as adaptive PageRank, maximal independent set, etc., and looking at the effectiveness of the two generalizable optimizations (direction-optimization and operand reuse), is another interesting direction.

## ACKNOWLEDGMENTS

We thank Yuechao Pan for valuable insight into BFS optimizations. We would like to acknowledge Scott McMillan for important feedback on early drafts of the paper. We appreciate funding from the National Science Foundation (Award # CCF-1629657), the DARPA XDATA program (US Army award W911QX-12-C-0059), and the DARPA HIVE program. For HIVE support, this material is based on research sponsored by Air Force Research Lab (AFRL) and the Defense Advanced Research Projects Agency (DARPA) under agreement number FA8650-18-2-7836.

This research was supported in part by the Applied Mathematics program of the DOE Office of Advanced Scientific Computing Research under Contract No. DE-AC02-05CH11231, and in part the Exascale Computing Project (17-SC-20-SC), a collaborative effort of the U.S. Department of Energy Office of Science and the National Nuclear Security Administration.

The U.S. Government is authorized to reproduce and distribute reprints for Governmental purposes notwithstanding any copyright notation thereon. The views and conclusions contained herein are those of the authors and should not be interpreted as necessarily representing the official policies or endorsements, either expressed or implied, of Air Force Research Lab (AFRL) and the Defense Advanced Research Projects Agency (DARPA) or the U.S. Government.